# Comment on 'Path Summation Formulation of the Master Equation'

The Green's function for a random walker on a lattice that starts at time $t_0=0$ at state $i$ and reaches at time $t$ state $j$, $G_{ji}(t)$, can be found by various approaches. For spatially invariant systems, a recursion relation finds the Green's function in Laplace-Fourier domains in terms of the basic input probability density functions (PDFs) [1]. For arbitrarily inhomogeneous bounded systems in discrete space, the solution for the Laplace transform Green's function inverts the matrix of transition rates with ($-s$) added to the diagonal elements ($s$ is the Laplace transform argument). For the general case, the elements of the inverted matrix are not known analytically. A different approach expresses the Green's function as a sum over all relevant paths, e. g. [2]. If one chooses to count and sum over paths, as was recently done when calculating the Green's function for one-dimensional arbitrarily inhomogeneous lattices with $L$ states [3], associating a combinatorial factor with each path is difficult. Only the product of the combinatorial factor and the path PDF in time is relevant in Green's function calculation. In higher dimensions, the combinatorial factor associated with a path becomes larger, and harder to compute.

A recent Letter claims to make a progress in summing over paths of a random walker in an arbitrarily inhomogeneous lattice of dimension D [4]. We believe this claim is unjustified; although Eq. (13) correctly expresses the Green's function as a sum over all possible relevant paths, one cannot plug in it Eq. (9) or Eq. (16) for a general case, and perform a summation, because combinatorial factors for paths are not given, or discussed, in [4]. (In this comment, we refer to the numbered equations that appear in [4].) (The only case for which one can calculate the Green's function by summing over paths without worrying about combinatorial factors is for a lattice with two states, where each path appears once.)

To illustrate our point, consider a one-dimensional four-state lattice with transition rate $k_{ji}$ connecting state $i$ to state $j$, and $K_i = \sum_j k_{ji}$. Say that the Laplace space Green's function $\overline{G}_{41}(s)$ is required. Using Eq. (13), $\overline{G}_{41}(s) = \sum_{\{x(n;s)\}} \overline{\Pi}_n[x(n;s)]$. How to find expression for the $\overline{\Pi}_n[x(n;s)]$s? For this, ideas given in [3] are used and briefly sketched below. First, the values $n=0, 1, 2$, do not contribute to the sum, because the random walker must perform at least 3 jumps to reach state 4 from state 1. (Similarly, all even values of $n$ do not contribute to the sum). For $n=3$, $x(3;s)$ is just the path of direct transitions, and it is denoted by, $x(3;s) = (1,2,3,4;s)$. Thus, $\overline{\Pi}_3[x(3;s)]$ is obtained using Eq. (9). For $n=5$, there are 3 distinct paths that contribute to the ensemble of paths $\{x(5;s)\}$, $\{x(5;s)\} = [(1,2,1,2,3,4;s), (1,2,3,2,3,4;s), (1,2,3,4,3,4;s)]$, and accordingly,

$$\overline{\Pi}_5[x(5;s)] = \overline{\Pi}_3[x(3;s)] \left( \frac{k_{21}k_{12}}{(s+K_1)(s+K_2)} + \frac{k_{32}k_{23}}{(s+K_2)(s+K_3)} + \frac{k_{43}k_{34}}{(s+K_3)(s+K_4)} \right).$$

The parenthesis above are defined as $\overline{h}(1,s;4)$. The next non-vanishing term in the series, $\overline{\Pi}_7[x(7;s)]$, is even more interesting, as it is built from degenerate paths; there are 2 paths with degeneracy 2 and additionally 4 paths with degeneracy 1. $\overline{\Pi}_7[x(7;s)]$ obeys the recursion relation, $\overline{\Pi}_7[x(7;s)] = \overline{\Pi}_5[x(5;s)]\overline{h}(1,s;4) - \overline{\Pi}_3[x(3;s)]\overline{h}(2,s;4)$, where $\overline{h}(2,s;4) = k_{21}k_{12}k_{43}k_{34} \prod_{i=1}^{4}(s+K_i)^{-1}$. This sort of recursion relation holds true for every odd $n$, leading to, $\overline{G}_{41}(s) = \overline{\Pi}_3[x(3;s)]/[1-\overline{h}(s,1;4)+\overline{h}(s,2;4)]$. Thus, $\overline{G}_{41}(s)$ cannot be approximated by the numerator, which represents the only path of direct transitions connecting states 1 and 4; it is the denominator, which is built from all paths with back transitions and their associate degeneracy, that gives the correct decaying rates and equilibrium probability. Finally, note that for the four-state chain, a complementary analysis shows that the number of paths associated with $\prod_n [x(n;t)]$ scales as $(1+\phi)^{n/2}$ where $\phi$ (=1.618…) is the golden ratio. This sort of complementary analysis can be useful when extending to higher dimensions the treatment given here and in [3] for 1$d$ for calculations of Green's functions and related approximations.

O. Flomenbom[1*], J. Klafter[2#], and R. J. Silbey[1§]

[1] *Department of Chemistry, Massachusetts Institute of Technology, Cambridge, MA 02139, USA*

[2] *School of Chemistry, Raymond & Beverly Sackler Faculty of Exact Sciences, Tel Aviv University, Ramat Aviv, Tel Aviv 69978, Israel*

[*] Electronic address: flomenbo@mit.edu
[#] Electronic address: klafter@post.tau.ac.il
[§] Electronic address: silbey@mit.edu

[1] E. W. Montroll & G. H. Weiss, J. Math. Phys. (N.Y.) **6**, 167 (1965); H. Scher & M. Lax, Phys. Rev. B **7**, 4491 (1973).

[2] R. D. Mattuck, *A guide to Feynman diagrams in the many-body problem,* McGraw-Hill International Book Company, USA, 1976.

[3] O. Flomenbom & J. Klafter, Phys. Rev. Lett. **95**, 098105 (2005).

[4] S. X. Sun, Phys. Rev. Lett. **96**, 210602 (2006).